\documentstyle[12pt,blois,psfig]{article}

\begin{document}
\heading{\bf Galaxy Clustering Determined From Numerical Cosmological Simulations.}

\author{ Adrian Jenkins$^{1}$, C.S. Frenk$^{1}$, F.R. Pearce$^{1}$, 
P.A. Thomas$^{2}$, J.M. Colberg$^{3}$, {S.D.M. White$^{3}$}, 
H.M.P. Couchman $^{4}$, J.A. Peacock $^{5}$, G. Efstathiou $^{6}$ and 
A.H. Nelson $^{7}$ $\phantom{xxxxxxxxxxxxxxxxxxx}$(The Virgo consortium)}
{$^{1}$ Dept Physics, South Road, Durham, DH1 3LE, UK}
{$^{2}$ CPES, University of  Sussex, Falmer, Brighton, BN1 9QH, UK}
{$^{3}$ MPA, Garching, Munich D-85740, Germany}
{$^{4}$ Dept of Astronomy, University of Western Ontario, London, 
Ontario, N6A 3K7,Canada}
{$^{5}$ Institute for Astronomy, Blackford Hill, Edinburgh, EH9 3HJ, UK}
{$^{6}$ Institute of Astronomy, Cambridge, CB2 OHA, UK}
{$^{7}$ Dept Physics and Astronomy, UWCC, Cardiff, CF2 3YB, UK}

\begin{bloisabstract}
   We have simulated the growth of structure in  two 100 Mpc boxes for
$\Lambda$CDM and SCDM universes. These N-body/SPH simulations include
a gaseous component which is able to cool radiatively. A fraction of
the gas cools into cold dense objects which we identify as galaxies.
In this article we give a preliminary analysis of the clustering
behaviour of these galaxies concentrating on the $\Lambda$CDM
model. We find a galaxy correlation function which is very close to a
power law and which evolves relatively little with redshift. The
pairwise dispersions of the galaxies are significantly lower than the
dark matter.  The $\Lambda$CDM model gives a
surprisingly good match to the observational determinations of the
galaxy correlation function and pairwise dispersions.  This article
should be read in conjunction with the article 'Cosmological Galaxy
Formation' by Frazer Pearce also in this volume.

\end{bloisabstract}

\section{Introduction}

  It is impossible at present numerically to model all the physical
processes which are involved in galaxy formation. Galaxies, as we see
them, are made of baryonic material, and at very least a gaseous
component must be included in any modelling.  In the hierarchical
galaxy formation scenario \cite{wr} dissipation by the baryons is also
crucial for galaxies to form.  However it is not clear that simply
including cooling of the gas is a sensible approach numerically as
this would appear to lead to catastrophic cooling in the early
universe.  The continual merging of dark halos leads to some shock
reheating of cold gas but it is not obvious that the cooling
catastrophy can be averted without help from some other heating
mechanism \cite{wf}.  The most likely heat source is that from star
forming regions, particularly supernovae explosions. It seems a
reasonable assumption that the energy injected back into the gas from
supernovae can act as a negative feedback mechanism and lead to some
kind of self-regulation of star formation.  This self-regulation is
expected to be at its most effective in the early universe when the
characteristic depth of the dark matter potential wells was much
lower than today.

  For our simulations we rely on an effect of numerical resolution to
mimic these feedback processes.  The finite mass resolution in the
simulation suppresses cooling in low mass halos and delays the onset
of the formation of the cold dense gas phase (defined at $T<10^4K$,
and $60\;000$ times the mean gas density) which we identify as
galaxies.  We do not include star formation in the simulation, but
assume that all the cold dense gas in the simulations should really be
in the form of stars.  The resolution is determined by the gas
particle mass. We selected a gas particle mass such that the amount of
gas in the cold dense phase at $z=0$ in the SCDM model provides a
reasonable match to the observed stellar mass density in the local
universe.

  With this  choice of gas particle mass, which equates to $2\times10^9
M_{sun}$, we are able to simulate a 100 Mpc cubed shaped region with
$128^3$ particles of both gas and dark matter.  This volume is large
enough to study clustering statistics on scales up to 10 Mpc with a 
sample of about 2000 galaxies.

 The simulations are started from a redshift of 50.
The models we have simulated are a $\Lambda$CDM model with
$\Omega_0 = 0.3$, $\Lambda_0 = 0.7$, $\sigma_8 = 0.9$, $h=0.7$,
$\Omega_b=0.03$ and an SCDM model with $\Omega = 1$, $\sigma_8 = 0.6$,
$h=0.5$, $\Omega_b=0.06$. The power spectra are the same as used for
the models in \cite{arj} except that we used a higher value of $\sigma_8$
for SCDM.

\begin{figure}
\centering
\hspace{0.in}\psfig{file=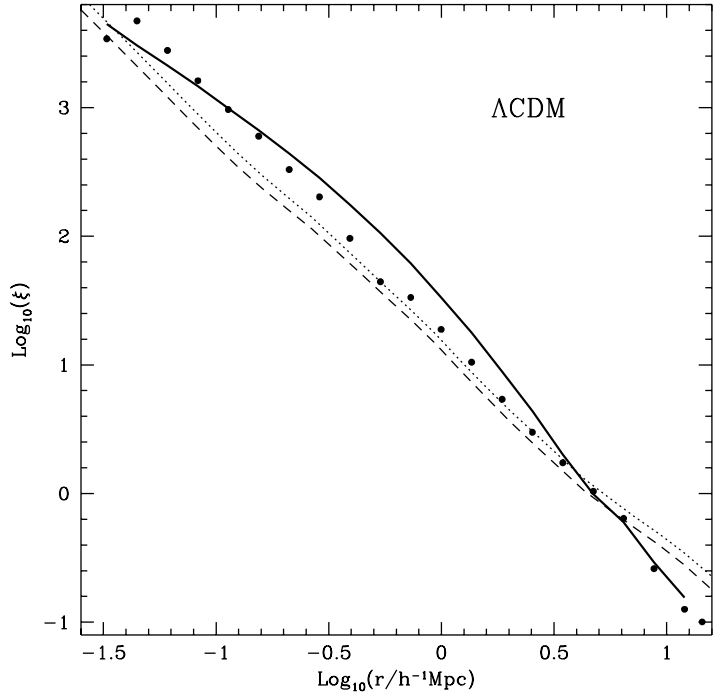,width=100mm}
\caption{The two-point correlation function of the ``galaxies'' for
the LCDM model (filled circles), compared to the mass correlation
function (solid line) and the observed galaxy real-space correlation
function (dashed/dotted - derived with different assumptions of
clustering evolution\cite{cmb})}
\label{fig1}
\end{figure}

\begin{figure}
\centering
\hspace{0.in}\psfig{file=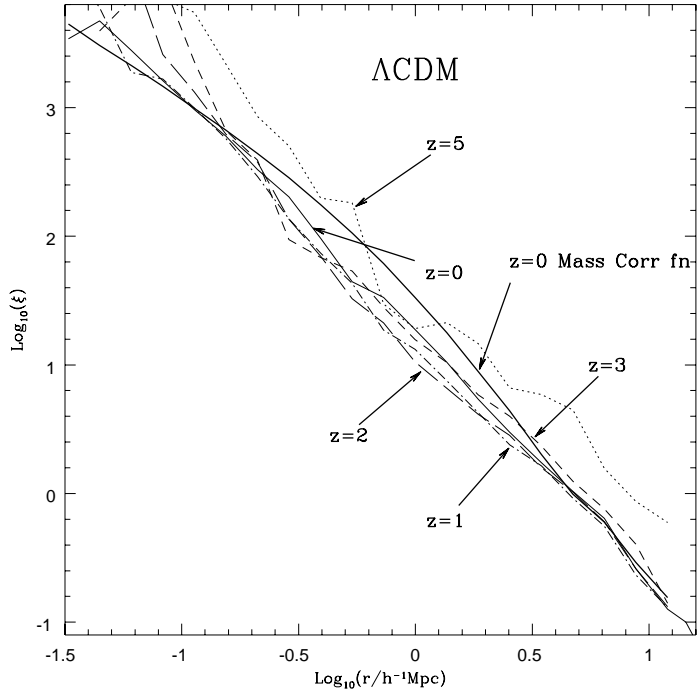,width=100mm}
\caption{Evolution of the galaxy correlation function from redshift 5 to
the present for the $\Lambda$CDM model. The bold line shows the $z=0$
mass correlation function.}
\label{fig2}
\end{figure}

\section{The clustering of galaxies}
 The galaxies in the simulation are identified using a
friends-of-friends group finder. The object catalog returned by 
the group finder is relatively insensitive to the linking length. We
chose a linking-length of 1/50 of the mean {\it physical} interparticle
separation at $z=0$.

 Figure 1 shows the two-point correlation functions for the simulated
galaxies (filled circles), and dark matter (solid line) in the
$\Lambda$CDM model. The observed galaxy correlation function
\cite{cmb} is shown as dashed and dotted lines.  The latter is from
\cite{cmb}. On large scales the galaxies and mass have essentially the
same correlation function. But on scales around $1Mpc/h$ the galaxy
distribution is anti-biased with respect to the mass.  The match to
the observed galaxy correlation function is pretty good - in
particular the simulated galaxy correlation function is close to a
power law unlike the mass correlation function. A power law galaxy
correlation function is also seen in the SCDM simulation.

\begin{figure}
\centering
\hspace{0.in}\psfig{file=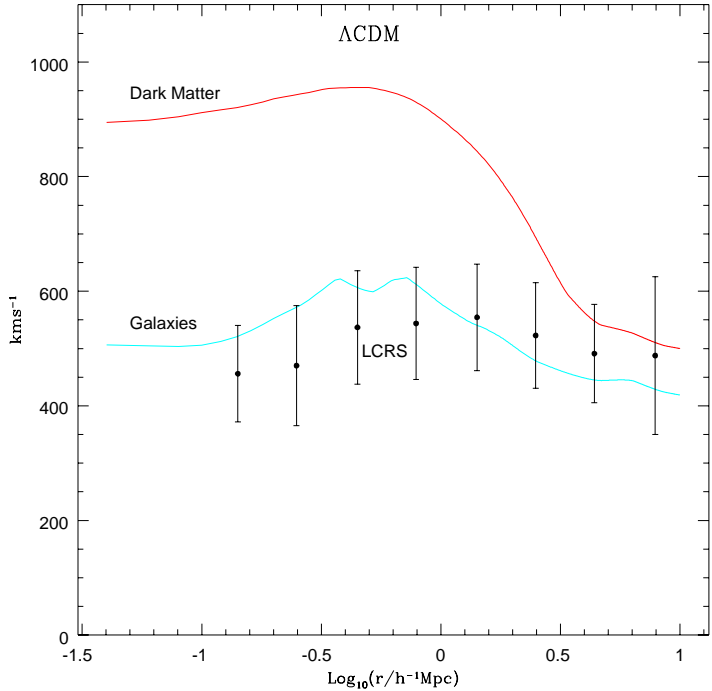,width=100mm}
\caption{Pairwise velocity dispersion for dark matter and galaxies. The
quantity plotted is the line-of-sight projection of the radial and 
transverse pairwise velocities (see \cite{arj} for details). Notice
the significant difference between the measured dispersion of the galaxies
and dark matter.}
\label{fig3}
\end{figure}

 Figure 2 shows the two-point correlation function of the galaxies in
the $\Lambda$CDM model at a range of epochs. The amplitude of the correlation
function has changed remarkably little from the present back until 
redshift of 3 even though the mass correlation function evolves linearly
by a factor 10. At redshift 5 the galaxy clustering strength is 
significantly greater than at the present.

Figure 3 shows the line-of-sight pairwise velocity dispersion
(explained in \cite{arj}) of the galaxies in the $\Lambda$CDM model
compared to the dark matter, and a determination from the LCRS
redshift survey \cite{jing}. The galaxies have a smaller dispersion
than the dark matter, something which is clearly necessary for the
viability of this cosmological model. We have found that if one
selects the nearest dark matter particle to each galaxy, then the
pairwise dispersion of these particles is essentially the same as the
galaxies.

\section{Summary}
  The clustering of the galaxies in the $\Lambda$CDM simulation is
surprisingly close to the observed galaxy correlation
function. Significantly in both $\Lambda$CDM and SCDM runs the galaxy
correlation function is much closer to a power law than the mass
correlation function.  There is little evolution in the galaxy
correlation function between redshift 3 and the present. For the
$\Lambda$CDM model the pairwise galaxy velocity dispersion is much
lower than that of the dark matter. A similar effect is seen in the pairwise
dispersions of galaxies in the SCDM model, but weaker.

\acknowledgements{The simulations described here were run at the 
Edinburgh Parallel Computing Centre by the Virgo Consortium. CSF
acknowledges a PPARC Senior Fellowship.}


%
\vfill
\end{document}